\documentclass[12pt]{article}
\usepackage{cite}
\usepackage{ulem}
\usepackage{color}
\usepackage{epsfig}
\usepackage{graphicx}
\usepackage{subfigure}
\usepackage{ulem}
\usepackage{amsfonts}
\usepackage{epsfig,bm}
\usepackage{graphicx}
\usepackage{comment}
\newcommand{\be}{\begin{equation}}
\newcommand{\ee}{\end{equation}}
\newcommand{\ba}{\begin{eqnarray}}
\newcommand{\ea}{\end{eqnarray}}
\newcommand{\bs}{\begin{scriptsize}}
\newcommand{\es}{\end{scriptsize}}
\newcommand{\nn}{\nonumber\\}
\begin{document}
\begin{titlepage}
\title{Scalar Moduli, Wall Crossing and Phenomenological Predictions}
\author{}
\date{Stefano Bellucci$^{b}$ \thanks{\noindent bellucci@lnf.infn.it} and
Bhupendra Nath Tiwari$^{b}$ \thanks{\noindent tiwari@lnf.infn.it}\\
\vspace{0.5cm}
$^{b}$INFN-Laboratori Nazionali di Frascati\\
Via E. Fermi 40, 00044 Frascati, Italy.}

\maketitle \abstract{We present the scalar moduli stabilization
from the perspective of the real intrinsic geometry. In this
paper, we describe the physical nature of the vacuum moduli
fluctuations of an arbitrary Fayet configuration. For finitely
many abelian scalar fields, we show that the framework of the real
intrinsic geometry investigates the mixing between the marginal
and threshold vacua. Interestingly, we find that the phenomena of
wall crossing and the search of the stable vacuum configurations,
pertaining to $D$-term and $F$-term scalar moduli, can be
accomplished for the abelian charges. For given vacuum expectation
values of the moduli scalars, we provide phenomenological aspects
of the vacuum fluctuations and phase transitions in the
supersymmetry breaking configurations.} \vspace{1.5cm}

\vspace{2mm} {\bf Keywords}: Wall Crossings, Intrinsic Geometry,
Moduli Configurations, Vacuum Stabilization, $D$-term potential,
$F$-term potential, Fayet Model, Scalar Phenomenologies.\\
\\
{\bf PACS}: String Theory: 11.25.-w; Gauge Theory: 47.20,Ky; Field
Theory: 11.27.+d; Intrinsic Geometry: 2.40.-Ky; Algebraic
Geometry: 2.10.-v; Statistical Fluctuation: 5.40.-a; Flow
Instability: 47.29.Ky
\end{titlepage}
\section{Introduction}
The physics of wall crossing \cite{wallcrossing} has received
considerable attention from the perspective of theoretical
developments, model building and phenomenological predictions.
Various prior supergravity models pertaining to the multiplet
structures have been developed in the literature
\cite{supergravitymultiplets}. In this concern, string theory
plays a crucial role in bringing out all possible spacetime
dimensions in a unified framework \cite{Witten}. Recently, there
have been a number of investigations involving $D$-brane
instantons in type-II compactifications \cite{Ralph} $D$-instanton
counting \cite{Morales} and corresponding microscopic properties.
The above first consideration leads to the notion of moduli
stabilization, while the second one leads to the localization of
an arbitrary $p$-form. Hereby, the hypermultiplet structure
\cite{Pioline} and the corresponding moduli space geometry play an
important role in understanding the phenomenon of wall crossing,
duality symmetries and twistor space properties. Various algebraic
properties of the wall crossing phenomenon opens up a new avenue
to learn modern issues of background scalar fields, viz., $(2,2)$
superconformal theories, $D$-terms, quasihomogeneous
superpotentials and spectral flow properties \cite{CecittiVafa}.

In sequel, black hole physics has revealed an understanding of
$AdS_5 \times S^5$ geometry \cite{Minwala} and the corresponding
microscopic aspects of a class of higher derivative corrected
black holes \cite{alphaprime}. The development does not stop here,
but it continues far beyond the expectation, viz. the overall
theme of microstate counting, $D$-brane instantons, and wall
crossing hints about the structure of the nonperturbative string
theory \cite{Larsen}. From the perspective of AdS/CFT
correspondence \cite{maldacena}, and hidden conformal field theory
\cite{Strominger}, this offers generalized thermodynamical
properties of BPS and non-BPS black holes, Hawking radiation,
symmetry breaking and the phenomenology of general black holes in
string theory. For the case of the two parameter Hawking radiating
black hole configurations \cite{BNTSB}, the constant amplitude
radiation and decreasing amplitude York model demonstrate that the
statistical fluctuations pertaining to horizon perturbations,
quantum gravity corrections and radiation flux can be globally
stabilized in specific frequency channels.

From the perspective of the real intrinsic Riemannian geometry,
the interesting cases arise for arbitrary charges of the gauged
supergravity and arbitrary component scalar moduli configurations.
A priori, we have analyzed the moduli stabilization properties of
the singlet and doublet complex scalar fields \cite{bntsbnov10},
which correspond to the pure marginal and threshold
configurations. Algebraically, the above consideration shows that
the BPS walls of moduli stability arise as certain polynomial
relations in the Fayet parameter. Here we will describe a much
more extensive class of exact, analytic stabilization properties
for the $D$-term and $F$-term moduli potentials. The present
solutions of intrinsic moduli stabilization have an arbitrary
number of scalar fields. Our solution may help to clarify many
moduli vacuum stabilization effects, which we have recently
introduced for the two and four components real scalar fields. As
per the notion of moduli configuration, our proposal may play an
important role in the exact phenomenological predictions of the
string theory moduli space configurations.

\section{Fayet Moduli Configurations}

The $D$-term moduli configuration may in general be described by
the potential \bs \ba V_D(\varphi_i) = (\sum_{i \in \Lambda} \ q_i
\ \varphi_i^{\dag} \varphi_i + b)^2, \ea \es where $b$ is the
Fayet term. Here we want to analyze ``walls of BPS stabilities" by
focusing our attention on the role of the intrinsic Riemannian
geometry. In order to keep arbitrary electric charges, we shall
consider the configuration in its most general form. It is worth
to recall that the marginal stability of the vacuum fluctuations
requires uniform sign of the electric charges, which must retain
the same sign as the Fayet term. Whilst, the threshold stability
requires relative signs of electric charges such that the $D$-term
can vanish, in order to satisfy the supersymmetry property of the
vacuum.  The main purpose of the present article is to illustrate
phenomenological properties of the moduli stabilization for such a
generic vacuum configuration. Let us first illustrate the
stabilization properties of the configurations with fewer
constituent moduli scalar fields. For a given complex modulus
$\varphi_i:= (x_1, x_2) \in U(1)$, the marginal stability of the
vacuum configuration is described by the following Fayet potential
\bs \ba V(x_1,x_2) := (x_1^2+x_2^2+b)^2. \ea \es

On the other hand, the corresponding threshold configuration with
two real scalar fields requires to consider two complex scalar
fields $\{ \varphi_1, \varphi_2\}$ in a special way. In order to
illustrate the role of the threshold type consideration, let us
first focus our attention on the moduli configuration involving
only two real scalars. In this case, it turns out that the general
vacuum configurations of the present interest can be defined as an
element of the moduli space, viz., $\{ \varphi_1, \varphi_2\} \in
\mathcal{M}_2$, such that the $D$-term potential vanishes in the
vacuum limit. In fact, it follows that the possible values of the
constituent fields are an element of $\mathcal{M}_2$, which in the
limit of extreme values of the fields, can be defined as the
following set \bs \ba \mathcal M_2&:=& \{ \{(x_1,0),(x_2,0)\},
\{(x_1,0),(0, x_2)\}, \nn && \{(0, x_1),(x_2,0)\}, \{(0, x_1),(0,
x_2)\} \}. \ea \es In order to ensure supersymmetry, we are
required to choose two alternating abelian charges, which in an
appropriate normalization correspond to the choice of $q_1= 1$ and
$q_2= -1$. Hereby, the threshold stability of the $D$-term moduli
configurations pertains to the following Fayet like potential \bs
\ba V(x_1,x_2) := (x_1^2-x_2^2+b)^2. \ea \es The two modulus $ \{
\varphi_1, \varphi_2 \}$ case involves four constituent real
scalar fields $(x_1, x_2, x_3, x_4)$. In this case, the $D$-term
potential, which corresponds to the marginal moduli configuration,
can be expressed as \bs \ba V(x_1,x_2,x_3,x_4) :=
(x_1^2+x_2^2+x_3^2+x_4^2+b)^2. \ea \es Correspondingly, the
$D$-term potential pertaining to the four constituent real scalar
fields, $x_a= (x_1, x_2,x_3, x_4)$, which describes the underlying
threshold moduli configuration, takes the following form \bs \ba
V(x_1,x_2,x_3,x_4) := (x_1^2+x_2^2-x_3^2-x_4^2+b)^2. \ea \es In
order to appreciate the intrinsic geometric version of the moduli
stabilization problem, we may consider a set of real scalar fields
$\{x_i\} \in \mathcal M$ such that the map $V: \mathcal M
\rightarrow R$ defines the real scalar potential of the complex
scalar moduli. Thus, we may geometrically analyze how the
transformations $b \mapsto -b$, $b \mapsto 0$ and $b \mapsto 1/b$
appear in the self pair correlation functions and the global
correlation volume on the vacuum moduli fluctuations. One of the
interesting transformation properties could be to examine how the
local and global vacuum correlation(s) behave under the above
dilation and translation of the Fayet parameter $b$.

Notice further that the reflection symmetry of the $D$-term
potential is required, due to the supersymmetry constraints.
Furthermore, the marginal configurations allow the decay of the
BPS configurations to non-BPS configurations. However, the
corresponding thresholds configurations do not. In particular, we
observe that the threshold configurations possess a nontrivial
vacuum moduli, for example $\mathcal M_2$, as mentioned for the
extreme limit of the two complex modulus configuration.
Practically, the perspective of the intrinsic geometry shows that
such a specification of the threshold configurations corresponds
to a particular orientation of the vacuum moduli with respect to
the line $b=0$, where the supersymmetry is restored.

\section{Intrinsic Geometry}

We will explore the moduli stabilization properties of the Fayet
like configurations for an arbitrary number of constituent scalar
fields. In general, let us first note that the consideration of
fractional charges and associated constituent gauge fields
motivates us to keep the charges to a set of general real values.
\subsection{$F$ Term Fluctuations}
To test the properties of our model, let us determine the local
and global intrinsic geometric properties corresponding to the
$F$-term potential. For given dilaton $\phi$ and axion $a$, the
$F$-term potential takes the form $V(x_1, x_2)= (x_1- x_2)^2$,
where the $x_1, x_2$ are functions of $\{ \phi, a \}$. For the
supersymmetry breaking configurations with $b\ne 0$ and
arbitrarily charged dilaton axion like configurations, the moduli
space potential can be cascaded as
\bs \ba V(x_1,x_2)&=& (q_1 x_1^2+q_2 x_2^2+b)^2, \ea \es where
$q_1, q_2$ are the abelian charges corresponding to the
constituent scalar fields $\{x_1, x_2\}$. In this case, we find
that the components of the metric tensor satisfy the following
quadratic polynomial relations
\bs \ba g_{x_1x_1} &=& 12 q_1^2 x_1^2+4 q_1 q_2 x_2^2+4 q_1 b, \nn
g_{x_1x_2} &=& 8 q_1 q_2 x_1 x_2, \nn g_{x_2x_2} &=& 4 q_1 q_2
x_1^2+ 12 q_2^2 x_2^2+ 4 q_2 b. \ea \es
From the perspective of the intrinsic geometry, we see that the
principle components of the intrinsic metric tensor are symmetric
quadratic polynomials in the real scalars, whereas the
off-diagonal component is a symmetric quadratic monomial. Hereby,
an analogous analysis may as well be performed easily for the
concerned global stability of the configuration. In particular, we
see that the determinant of the metric tensor takes a well-defined
positive-definite quadratic polynomial form
\bs \ba \Vert g \Vert= 16 q_1 q_2 (q_1 x_1^2+q_2 x_2^2+b) (b+3 q_2
x_2^2+3 q_1 x_1^2). \ea \es
From the definition of the Gaussian fluctuations, we find that the
underlying vacuum moduli configuration has the following invariant
scalar curvature
\bs \ba R &=& -b \frac{q_1 x_1^2+q_2 x_2^2}{(q_1 x_1^2+q_2
x_2^2+b)^2(b+3 q_2 x_2^2+3 q_1 x_1^2)^2}. \ea \es
For a set of given charges, we see that the scalar curvature
vanishes, for the vanishing value of the Fayet parameter. In this
case, the determinant of the metric tensor remains a nonzero
constant, for the vanishing value of the Fayet parameter. The
present investigation demonstrates that the $F$ term scalar moduli
configurations are statistically interacting, as long as the
vacuum possesses a nonzero Fayet parameter and both the vacuum
expectation values, viz., $\{x_1,x_2\}$, do not vanish
simultaneously. For all values of the Fayet parameter, it is
interesting to notice that the threshold configurations, which are
defined as a pair of scalar fields with alternating abelian
charges, correspond to a noninteracting statistical basis,
whenever the constituent scalar moduli fields take an equal
absolute value in the vacuum.

\subsection{$D$ Term Fluctuations}
Let us now consider a realistic case of the vacuum moduli
fluctuation. For arbitrary abelian charges $\{q_1, q_2, q_3, q_4
\} $, a general complex doublet, when the underlying theory has
been broken to the $U(1) \times U(1)$, involves the following
Fayet potential
\bs \ba V(x_1,x_2,x_3,x_4)&=& (\sum_{i=1}^4 q_i x_i^2+b)^2. \ea
\es
Herewith, we observe that the local moduli pair correlations
reduce to the following expressions
\bs \ba g_{x_ix_i}&=& 12 q_i x_i^2+4 q_i \sum_{j \neq i } q_j
x_j^2 +4 q_i b, \ i= 1,2,3,4 \nn g_{x_ix_j} &=& 8 q_iq_j x_ix_j, \
\forall i \neq j, \ i,j= 1,2,3,4. \ea \es
As mentioned before, in this case, we find further that the
principle components of the metric tensor are polynomials in the
scalar fields, while, the off-diagonal terms are monomials in the
concerned scalar fields. In order to analyze the internal
stability of the configuration, we need to examine the positivity
of the principle minors of the metric tensor. In this case, we
find that the planar and hyper planar minors are given by the
following polynomials
\bs \ba g_2 &=& 16 q_1 q_2 (\sum_{i=1}^4 q_i x_i^2+b)
(b+3\sum_{i=1}^2 q_i x_i^2+ \sum_{i=3}^4 q_i x_i^2), \nn
g_3 &=& 64 q_1 q_2 q_3 (\sum_{i=1}^4 q_i x_i^2+b)^2 (b+3
\sum_{i=1}^3q_i x_i^2+q_4 x_4^2). \ea \es
The pattern of the polynomial invariance continues and the
corresponding determinant of the metric tensor satisfies the
following polynomial expression
\bs \ba \Vert g \Vert&=& 256 q_1 q_2 q_3 q_4 (\sum_{i=1}^4 q_i
x_i^2+b)^3 (3 \sum_{i=1}^4 q_i x_i^2+b). \ea \es
An analogous analysis may as well be performed further for the
computation of the concerned global intrinsic geometric invariant
quantity. In particular, we see that the underlying Ricci scalar
of the vacuum moduli fluctuations is given by
\bs \ba R &=& -\frac{9}{2} \frac{ \sum_{i=1}^4 q_i x_i^2 } {(3
\sum_{i=1}^4 q_i x_i^2+b )^2 (\sum_{i=1}^4 q_i x_i^2+b)}. \ea \es
In this case, we find that the vacuum moduli configurations,
defined by the real scalar fields $\{x_1,x_2,x_3,x_4\}$,
correspond to an interacting statistical syatem for all values of
the Fayet parameter. It is worth mentioning that the statistical
interactions continue to persist, even for the case when the Fayet
parameter vanishes. Further, the global statistical interactions
exist in the limiting marginal moduli system with an equal value
of the constituent scalar fields.

\begin{figure*}
\includegraphics[height=.3\textheight,angle=0]{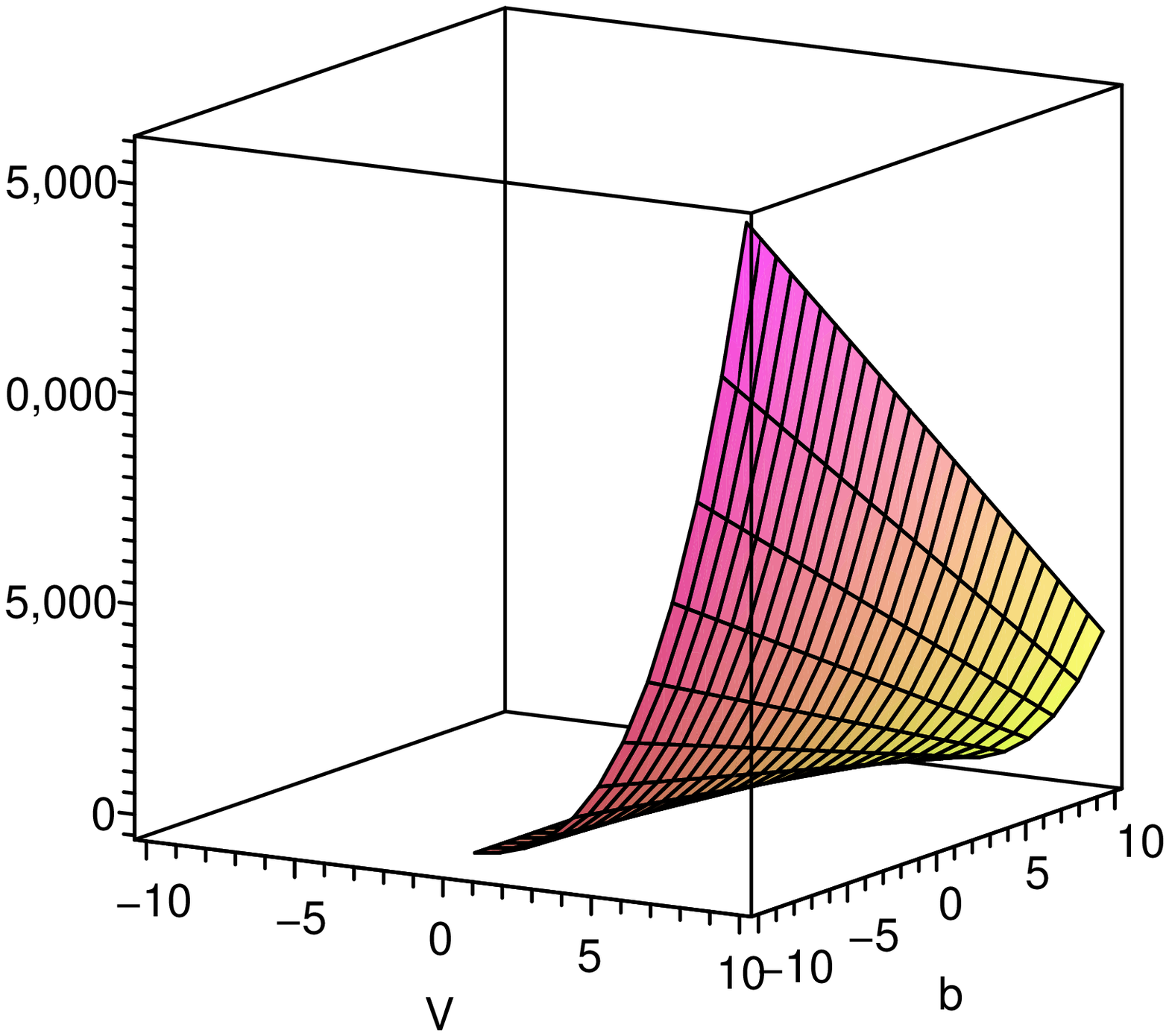}
\includegraphics[height=.3\textheight,angle=0]{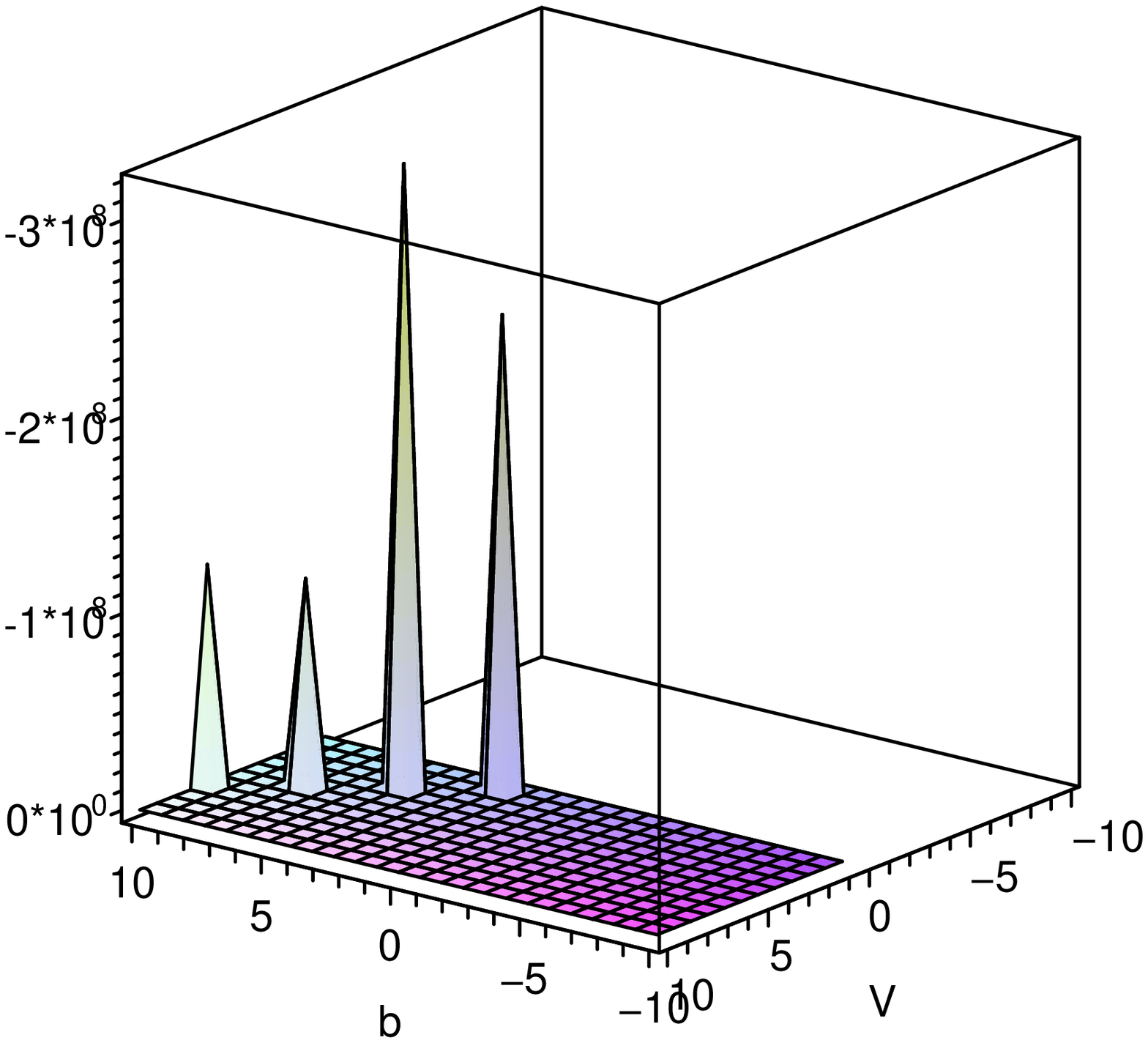}
\caption{The determinant of the metric and scalar curvature as a
function of the potential energy $V$ and Fayet parameter $b$ for
the six component real vacuum moduli configuration. As per this
prescription, note that we can measure the vacuum fluctuations of
an arbitrary finite component scalar moduli configuration.}
\label{corr}
\end{figure*}

This shows a clear cut distinction between the global vacuum
statistical interactions of the $F$ term moduli and $D$ term
moduli configurations. It is observed that the marginal moduli
configurations preserved the nature of intrinsic geometric
invariants. Specifically, the determinant of the metric tensor
remains positive for nontrivial vacua with a given Fayet
parameter. However, the threshold vacua change their statistical
behavior by a translation in the vacuum expectation values of the
constituent scalar fields. For a set of given abelian charges, we
find that the global nature of the two and four real scalar moduli
marginal and threshold configurations is characterized as
\bs \ba R_{mar}(x_i)\vert_{x_i= x} &\ne& 0,\ \
R_{thr}(x)\vert_{x_i= x} = 0, i= 1,2 \nn R_{mar}(x_i)\vert_{x_i=
x} &\ne& 0,\ \ R_{thr}(x_i)\vert_{x_i= x} \ne 0, i= 1,2,3,4. \ea
\es
We have generalized the above computation for the six and higher
number of real scalars. In fact, we have obtained the exact
expressions for the components of the metric tensor, principle
minors, determinant of the metric tensor and the underlying scalar
curvature of the fluctuating configurations. As per this
evaluation, the global correlation properties of the six moduli
configuration are shown in Fig.(\ref{corr}). Under the vacuum
fluctuations, the first figure describes the ensemble stability of
the vacuum configuration and the second figure describes the
corresponding phase space stability of the vacuum moduli
configuration.

\section{Scalar Phenomenologies}
Finally, let us describe the mixing of finitely many real scalar
fields. For a set of given abelian charges, the statistical
fluctuation of the most general Fayet like configuration involves
an ensemble of vacuum moduli fields. As mentioned in the foregoing
case, we now consider an extension of the statistical fluctuations
pertaining to the phenomenology of the Fayet models, with a set of
finitely many constituent real scalar fields. The picture of the
marginal and threshold configurations continues to hold for an
ensemble of the constituent scalar fields. For $q_i \in R$, let
$V(\{x_i\})= (\sum_{i \in \Lambda} q_i x_i^2)^2, \Lambda \subset
Z$ be the Fayet potential. Then, the global vacuum correlation
volume, when considered as the scalar curvature of the
corresponding statistical configuration, is described as
\bs \ba R(\{x_i\}= (const). \frac{\sum_{i \in \Lambda} q_i
x_i^2}{(b+ 3\sum_{i \in \Lambda} q_i x_i^2)^2}
\frac{(\tilde{\beta} b+ \tilde{\alpha} \sum_{i \in \Lambda} q_i
x_i^2)}{\Vert g \Vert_{linear}}, \ea \es
where $\Vert g \Vert_{linear}$ is a $C^{\infty}$ smooth linear map
in the $V: \mathcal M \rightarrow R$. In this model, we find that
the parameters $\{\tilde{\alpha}, \tilde{\beta} \}$ can be
subsequently determined from the phenomenological considerations
of supersymmetry breaking to $U(1)^r$. Hereby, we observe that the
above claim describes an arbitrary mixing of threshold and
marginal configurations, viz., uniform sign of the weight $q_i$
for the marginal configurations and alternating sign for the
threshold configurations.

Following the foregoing pattern, we find that the $D$ term and $F$
term configurations can completely be determined as a straight
line plot between the square root of the Fayet potential and the
Fayet parameter. Herewith, from the principle of the mathematical
induction, we find further justifications for an increasing number
of the constituent scalar fields. In general, let there be $m$
real scalar moduli. Then, as per the pattern observed in the
foregoing configurations, we observe that the corresponding
intermediate principle minors $p_n$ and the determinant of the
metric tensor take the following from
\bs \ba p_n&=& (\prod_{i=1}^n 4^i q_i) V(\{x_i\})^{\frac{m-1}{2}}
(\sqrt{V} + 2 \sum_{i=1}^n q_i x_i^2),  \nn \Vert g \Vert &=& p_m=
(\prod_{i=1}^n 4^i q_i) V(\{x_i\})^{\frac{m-1}{2}} (3 \sqrt{V}-
2b), \ea \es
where we have taken an account of the fact that arbitrary Fayet
potential $V(\{x_i\})$ admits a linear combination in the square
of the constituent scalar moduli, viz., $\sqrt{V}-b= \sum_{i \in
\Lambda} q_i x_i^2$. In this convention, we find the following
expression for the scalar curvature \bs \ba R= - (const)
\frac{(\sqrt{V}-b)}{V (3 \sqrt{V}- 2b)^2} (\alpha \sqrt{V}+ \beta
b). \ea \es Based on the above observation, we conjecture $\forall
i \in \Lambda$ that $\alpha, \beta \in Z$. A physical
justification is provided as follows. In doing so, we can easily
enlist the specific configurations while verifying the above
conjecture, and thus a classification of the scalar moduli
configurations. In particular, we find that the vacuum statistical
fluctuations can be described by specifying the parameters of the
underlying scalar moduli configurations, viz., (i) 2 real scalars:
$\alpha= 0, \beta=1$, (ii) 4 real scalars: $\alpha=1, \beta=0$ and
(iii) 6 real scalars: $\alpha=3, \beta=-1$.
For the supersymmetric models broken down to $U(1)^r$, we hereby
find that the lines of the vacuum phase transitions are $V=0$ and
$V= 4b^2/9$. From the perspective of the local and global mixing,
we observe further that the global correlation volume, viz. the
scalar curvature, of finitely many scalar moduli configurations
can be expressed as \bs \ba R= - \frac{const}{\Vert g \Vert^2}
V^{2 \gamma}(\sqrt{V}- b) (\alpha \sqrt{V}+ \beta b). \ea \es This
leads to an observation of the fact that the phenomenology of the
scalar moduli correlations arises with $\gamma= k-1$ and $m=2k $.
In order to experimentally determine the admissible numerical
values of the model parameters $\{ \alpha, \beta \}$, let us
define $\delta:= -(const) \frac{(\sqrt{V}-b)}{V (3 \sqrt{V}-
2b)^2}$. Thus, for a given vacuum scalar curvature $R$, the Fayet
potential and the Fayet parameter satisfy the following linear
relation \bs \ba \sqrt{V}= - \frac{\beta}{\alpha} b+
\frac{R}{\delta \alpha}. \ea \es
Experimentally, the parameters $\{\alpha, \beta \}$ can be
determined from the slope angle $ \theta=
-tan^{-1}(\frac{\beta}{\alpha})$ and its intercept
$c=\frac{R}{\delta \alpha}$ on the $b$-axis. Specifically, we see
that the curve of interaction, as per the characterization of the
above straight line, passes from the first and third quadrants.
Physically, the constant $\{\alpha, \beta \}$ can be fixed by
scaling the scalar potential at a different Fayet parameter. In
fact, it follows that our consideration involves only the globally
invariant determinant of the metric tensor in the globally
invariant scalar curvature, and thus only the correlation
dimensions of the vacuum moduli. For a well-defined equilibrium
configuration, we find that the underlying statistical system
becomes noninteracting for the following case of (i) vanishing
vacuum expectation values of the constituent scalar fields, (ii)
the square root of the sum of vacuum expectation values of the
scalar fields becomes equal to the value of the Fayet parameter
and (iii) the parameters of vacuum configuration vanish
identically, viz., $\alpha=0, \beta=0$. This provides an
excitement for phenomenological aspects of the vacuum fluctuations
in the $D$ and $F$ scalar moduli configurations.

\section*{Acknowledgement}
Supported in part by the European Research Council grant
n.~226455, \textit{``SUPERSYMMETRY, QUANTUM GRAVITY AND GAUGE
FIELDS (SUPERFIELDS)"}. 
%


\begin{thebibliography}{99}
%
\bibitem{wallcrossing} E. Witten, JHEP {\bf 04} (2002) 012;
A. Sen, JHEP {\bf 0705} (2007) 039; A. Sen, JHEP {\bf 07} (2008)
078; A. Mukherjee, S. Mukhi, R. Nigam, Mod. Phys. Lett. A {\bf 24}
(2009) 1507-1515; D. Gaiotto, G. W. Moore, and A. Neitzke, Commun.
Math. Phys. {\bf 299} (2010) 163-224; E. Diaconescu, G. W. Moore,
{\tt arXiv:0706.3193v1 [hep-th]}; F. Denef, JHEP {\bf 10} (2002)
023; F. Denef, JHEP {\bf 08} (2000) 050; D. Joyce, Adv. Math. 217
(2008), no. 1, 125-204; N. Nekrasov, E. Witten, arXiv:1002.0888v2
[hep-th]; N. A. Nekrasov, Adv. Theor. Math. Phys. {\bf 7} (2004)
831-864.

\bibitem{supergravitymultiplets}
S. Ferrara, A. Gnecchi, A. Marrani, Phys. Rev. D {\bf 78} (2008)
065003; R. D'Auria, S. Ferrara, M. Trigiante. Nucl. Phys. B {\bf
780} (2007) 28-39; S. Bellucci, S. Ferrara, A. Marrani, A.
Yeranyan, Phys. Rev. D {\bf 77} (2008) 085027; G. Gibbons, R.
Kallosh, B. Kol, Phys. Rev. Lett. {\bf 77} (1996) 4992-4995; D.
Green, E. Silverstein, D. Starr, Phys. Rev. D {\bf 74} (2006); G.
Villadoro, F. Zwirner, JHEP {\bf 0603} (2006) 087.

\bibitem{Witten} E. Witten, Nucl. Phys. B {\bf 443} (1995) 85-126

\bibitem{Ralph} R. Blumenhagen, M. Cvetic, S. Kachru, T. Weigand,
Ann. Rev. Nucl. Part. Sci. {\bf 59} (2009) 269-296; R.
Blumenhagen, S. Moster, E. Plauschinn, Phys. Rev. D {\bf 78}
(2008) 066008.

\bibitem{Morales} N. Seiberg and E. Witten, Nucl. Phys. B {\bf 431}, (1994) 19;
D. Bellisai, F. Fucito, A. Tanzini, G. Travaglino, Phys. Lett. B
{\bf 480} (2000) 365-372; U. Bruzzo, F. Fucito, J. F. Morales, A.
Tanzini, JHEP {\bf 0305} (2003) 054; M. Bianchi, F. Fucito, J. F.
Morales, JHEP {\bf 0908} (2009) 040.

\bibitem{Pioline} B. Pioline, Class. Quant. Grav. {\bf 23} (2006)
S981; M. Gunaydin, A. Neitzke, B. Pioline, A. Waldron, JHEP {\bf
0709} (2007) 056; J. Manschot, B. Pioline, A. Sen,
arXiv:1011.1258v2 [hep-th].

\bibitem{CecittiVafa} S. Cecotti, Nuc. Phys. B {\bf 355}, 3,
27 (1991), 755-775; S. Cecotti and C. Vafa, Commun. Math. Phys.
{\bf 158} (1993) 569-644; arXiv:0910.2615v2 [hep-th];
arXiv:1002.3638v1 [hep-th]; J. J. Heckman, C. Vafa, JHEP {\bf
0709} (2007) 011.

\bibitem{Minwala} S. Bhattacharyya, S. Lahiri, R. Loganayagam,
S. Minwalla, JHEP {\bf 0809} (2008) 054; S. Bhattacharyya, S.
Minwalla, JHEP {\bf 0909} (2009) 034; P. Basu, J. Bhattacharya, S.
Bhattacharyya, R. Loganayagam, S. Minwalla, V. Umesh, JHEP {\bf
1010} (2010) 045; S. Bhattacharyya, S. Minwalla, K. Papadodimas,
arXiv:1005.1287v1 [hep-th].

\bibitem{alphaprime} R. M. Wald, Phys. Rev. D {\bf 48} (1993)
3427-3431; R. M. Wald, Living Rev. Rel. {\bf 4} (2001) 6; V. Iyer,
R. M. Wald, Phys. Rev. D {\bf 50} (1994) 846-864; S. E. Gralla, R.
M. Wald, Class. Quant. Grav. {\bf 25} (2008) 205009; A. Sen, JHEP
{\bf 0509} (2005) 038; A. Sen, Gen. Rel. Grav. {\bf 40} (2008)
2249-2431; A. Sen, Int. J. Mod. Phys. A {\bf 24} (2009) 4225-4244;
A. Sen, JHEP {\bf 08} (2009) 068; S. Murthy and B. Pioline, JHEP
{\bf 09} (2009) 022; A. Dabholkar, J. Gomes, S. Murthy, A. Sen,
arXiv:1009.3226v1 [hep-th].

\bibitem{Larsen} A. Castro, D. Grumiller, F. Larsen, R. McNees,
JHEP {\bf 0811} (2008) 052; E. Gimon, F. Larsen, J. Simon, JHEP
{\bf 0907} (2009) 052; A. Castro, F. Larsen, JHEP {\bf 0912}
(2009) 037; M. Cvetic, F. Larsen, JHEP {\bf 0909} (2009) 088; A.
Castro, C. Keeler, F. Larsen, arXiv:1004.0554v1 [hep-th].

\bibitem{maldacena} J. M. Maldacena, Adv. Theor. Math. Phys. {\bf 2}
(1998) 231-252; O. Aharony, S. S. Gubser, J. M. Maldacena, H.
Ooguri, Y. Oz, Phys. Rept. {\bf 323} (2000) 183-386; I. R.
Klebanov, E. Witten, Nucl. Phys. B {\bf 556} (1999) 89-114; E.
Silverstein, E. Witten, Nucl. Phys. B {\bf 444} (1995) 161-190.

\bibitem{Strominger} A. Castro, A. Maloney, A. Strominger,
arXiv:1004.0996v1 [hep-th]; M. Guica, A. Strominger,
arXiv:1009.5039v1 [hep-th]; M. Guica, T. Hartman, W. Song, A.
Strominger, arXiv:0809.4266v1 [hep-th]; T. Hartman, K. Murata, T.
Nishioka, A. Strominger, JHEP {\bf 0904} (2009) 019; A. Maloney,
W.Song, A. Strominger, Phys. Rev. D {\bf 81} (2010) 064007; T.
Hartman, W. Song, A. Strominger, arXiv:0912.4265v1 [hep-th].

\bibitem{BNTSB} S. Bellucci, B. N. Tiwari, JHEP {\bf 1011}
(2010) 030.

\bibitem{bntsbnov10} S. Bellucci, B. N. Tiwari, arXiv:1011.3406
[hep-th].
%
\end{thebibliography}
\end{document}